# Helical Close Packings of Ideal Ropes

Sylwester Przybył [1], Piotr Pierański [1,2]

[1] Poznań University of Technology,
Piotrowo 3, 60 965 Poznań, Poland
e-mail: pieransk@man.poznan.pl
phone: +48 606 81 40 46

[2] Institute of Molecular Physics,
Smoluchowskiego 17, 60 159 Poznan, Poland

**ABSTRACT**

Closely packed conformations of helices formed on the ideal rope are considered. The pitch versus radius relations which define a closely packed helix are determined. The relations stem from the turn-to-turn distance and curvature limiting conditions. Plots of the relations are shown to cross each other. The physical sense of the crossing point is discussed.

## 1. Introduction

The phenomenon of coiling is observed in various biological and physical systems such as the tendrils of climbing plants [1], one-dimensional filaments of bacteria [2] or cylindrical stacks of phospholipid membranes interacting with an amphiphilic polymers [3]. In some cases the phenomenon occurs in conditions in which the helical structures created by coiling become closely packed. A single, closely packed helix is one of them. In a different context its formation was studied in Monte-Carlo simulations by Maritan et. al. [4]. Below we present a simple analytical arguments leading to the determination of the parameters of the optimal closely packed helix.

Take a piece of a rope of diameter $D$ and try to arrange it into a right-handed helix, see Fig.1, described parametrically by the set of equations:

$$x = -r\sin(\xi)$$
$$y = r\cos(\xi) \qquad (1)$$
$$z = \frac{P}{2\pi}\xi$$

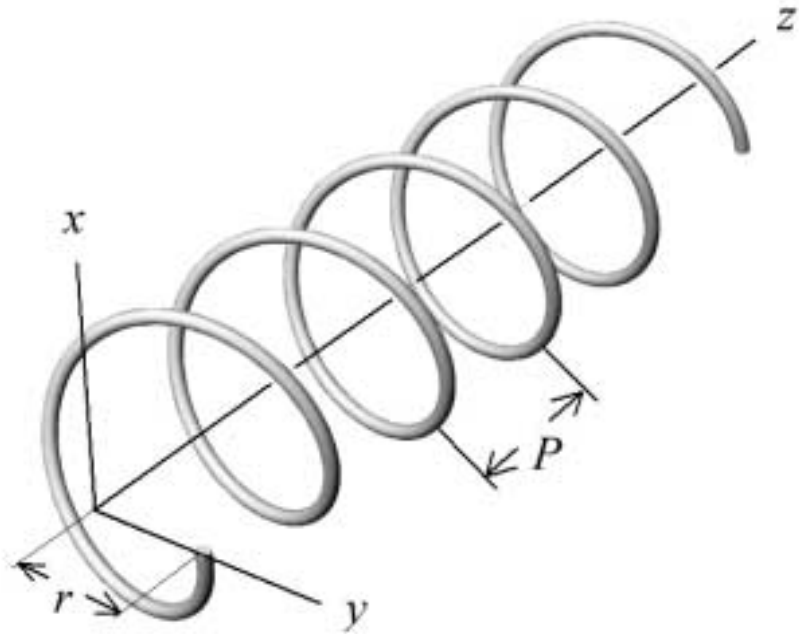

Fig. 1 The radius and pitch of a helix defined by eq. (1)

The helix is well defined if its radius $r$ and pitch $P$ are specified. As easy to check experimentally, when the helix is formed on a real rope, not all values of $r$ and $P$ are accessible. Being material, the rope cannot be arranged into shapes, which violate its self-impenetrability. If, for instance, one chooses to form a helix with $r=2D$, its pitch cannot be made smaller than about 1.003 $D$. This is the value at which the consecutive turns of the helix become closely packed. For a smaller pitch overlaps would occur. See Fig. 2.

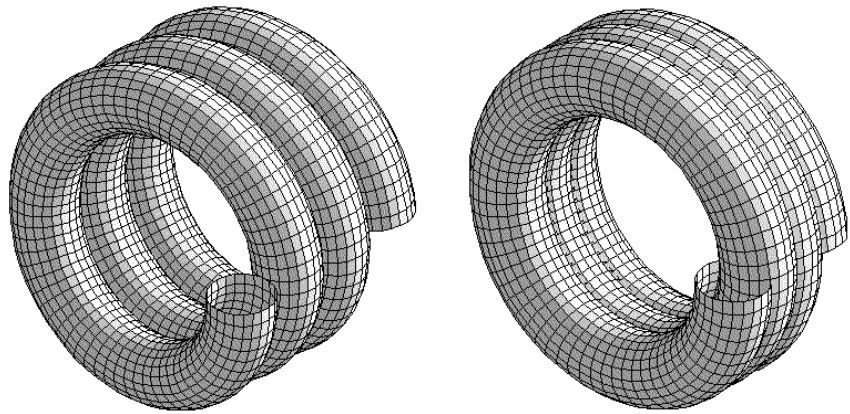

Fig. 2  When at a given radius the pitch of the rope helix is too small overlaps appear.
$r=2$; $P=1.003$ on the left and $P=0.5$ on the right.

A general question arises: *which are the limits for the radius and pitch values of a helix formed with a rope of the diameter D?*
The question was posed some time ago in discussions connected with the problem of ideal knots [5]. Calculations we present here were directly stimulated by recent paper by Maritan et al. who performed Monte Carlo simulations of the rope confined within a box. As indicated by the authors, the pitch to radius ratio of the optimal helix they discovered matches very well the value of the ratio found in the α-helix discovered by nature in the evolution processes. Possible implications of their results were discussed by Stasiak and Maddocks [6].

The symbolic algebra and numerical techniques we present below are analogous to those we applied considering the problem of the close packings of two ropes twisted together [7].

**2. Ideal rope**

Considerations presented below are valid for the so-called ideal ropes, i.e. ropes, which, from a physical point of view are completely flexible, but at the same time perfectly hard. Assume that the axis of such a rope is shaped into a smooth curve *C*. At any point of the curve its tangent vector ***t*** is defined. The rope is ideal, if each of its sections, perpendicular to the tangent vector ***t***, makes a circle of diameter *D*. None of the circles overlaps with any other of them.

Let the ideal rope be shaped into a closely packed helix *H* of a radius *r>D*. To understand better, what we mean by the „closely packed helix", we may imagine that the rope is wound as tightly as possible on a cylinder of diameter (*2r-D*).

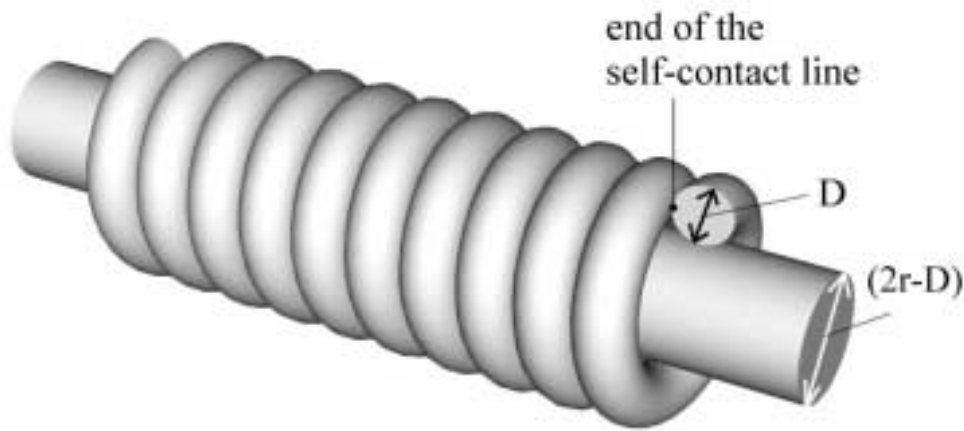

Fig.3 Rope a diameter *D* wound as tightly as possible
on a cylinder of diameter (*2r-D*).

In such conditions *the consecutive turns of the helix remain at the smallest possible distance equal D*. The points, at which the closely packed rope stays in touch with itself, are located on a helix *H'* of a radius *r'<r* (the pitch of *H'* remains the same as that of *H*). Now, let us remove the inner cylinder and try to make *r* smaller and smaller keeping all time the helix closely packed. As *r* goes down, the pitch *P* goes up. Below we determine the relation $P_{CP1}(r)$ which in the minimum turn-to-turn distance condition binds *r* and *P*. Experiments prove that at a certain value of *r*, the path determined by the $P_{CP1}(r)$ relation is left and the consecutive stage of the squeezing process is governed by a different limiting condition: *the local curvature κ of the helix cannot be larger than 2/D*. This new condition binds *P* and *r* in a different manner. We shall also find its shape $P_{CP2}$ $(r)$. At the point, at which both relations cross, the closely packed helix has a special, optimal shape discussed by Maritan et.al. [4].

## 3. Closely packed helix limited by its doubly critical self distance

Consider the ideal rope shaped into a helix $H$ whose consecutive turns touch each other. Which, for a given $r$, should be the pitch $P$ of the helix, to keep its turns closely packed? We shall answer the question.

Consecutive turns of the helix touch each other if the minimum of the distance from any point $P_1$ of the helix to the points located in the beginning of the next turn is equal $D$. Let $P_2$ be the point at which this minimum distance is reached. Let $\boldsymbol{t}_1$ and $\boldsymbol{t}_2$ be the vectors tangent of $H$ at $P_1$ and $P_2$, respectively. Obviously, in such a situation, the $\overrightarrow{P_1P_2}$ vector is perpendicular both to $\boldsymbol{t}_1$ and $\boldsymbol{t}_2$. $\overrightarrow{P_1P_2}$ belongs both to the plane $\Sigma_1$ located at $P_1$ and perpendicular to $\boldsymbol{t}_1$ and to the plane $\Sigma_2$ located at $P_2$ and perpendicular to $\boldsymbol{t}_2$. Thus, $\overrightarrow{P_1P_2}$ belongs to the line along which $\Sigma_1$ and $\Sigma_2$ cross. Let $P_1$ located at $(x_1, y_1, z_1)$ be given; let it be the point of $H$ defined by $\xi_l = 0$:

$$\begin{aligned} x_1 &= 0 \\ y_1 &= r \qquad (2) \\ z_1 &= 0 \end{aligned}$$

Components of the tangent vector $\boldsymbol{t}_1$ located at $P_1$ are equal:

$$\begin{aligned} t_{1x} &= -r \\ t_{1y} &= 0 \qquad (3) \\ t_{1z} &= \frac{P}{2\pi} \end{aligned}$$

Consequently, the $\Sigma_1$ plane going through $P_1$ and perpendicular to $\boldsymbol{t}_1$ is defined by the equation:

$$rx - \frac{P}{2\pi} z = 0, \qquad (4)$$

Let

$$\begin{aligned} x_2 &= -r\sin(\xi) \\ y_2 &= r\cos(\xi) \qquad (5) \\ z_2 &= \frac{P}{2\pi} \xi \end{aligned}$$

be the coordinates of the $P_2$ point located in the vicinity of the next turn, i.e. at the $\xi$ values close to $2\pi$. The point must belong to the $\Sigma_1$ plane. Consequently its coordinates given by (5) must fulfil equation (4), what gives:

$$r^2 \sin(\xi) + \frac{P^2}{4\pi^2}\xi = 0 \qquad (6)$$

The square of distance between $P_1$ and $P_2$ should be equal $D^2$ what results in equation:

$$2r^2 - 2r^2 \cos(\xi) + \left(\frac{P}{2\pi}\right)^2 \xi^2 = D^2 \qquad (7)$$

Solving equations (6) and (7) for $r$ and $P$ gives a set of formulas:

$$P = 2\pi D \sqrt{\frac{\sin(\xi)}{2\xi[\cos(\xi)-1] + \xi^2 \sin(\xi)}}$$
$$r = D \sqrt{\frac{\xi}{2\xi[1-\cos(\xi)] - \xi^2 \sin(\xi)}} \qquad (8)$$

which in a parametric manner describe the relation $P_{CP1}(r)$ between $P$ and $r$ which must be fulfilled by helices whose consecutive turns are closely packed. (Notice that $\xi$ becomes here a free parameter which serves to describe the shape of the $P_{CP1}(r)$ relation.) The shapes of the $P(\xi)$ and $r(\xi)$ functions are shown in Fig. 4, where we present them in the potentially interesting interval of $\xi \in (\pi, 2\pi)$.

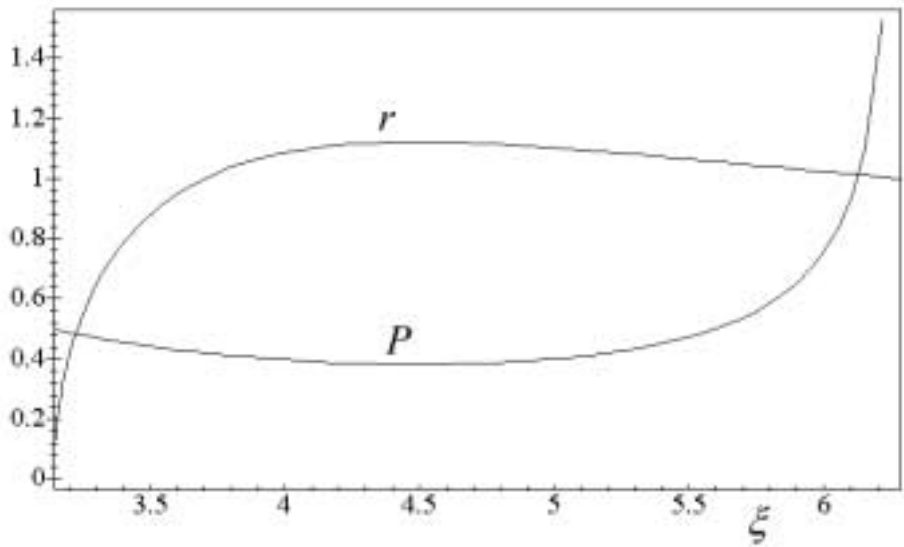

Fig.4   Shapes the $P(\xi)$ and $r(\xi)$ functions given by (7). $D$=1.

The shape of the $P_{CP1}(r)$ relation obtained within the parametric plot is shown in Fig.5.

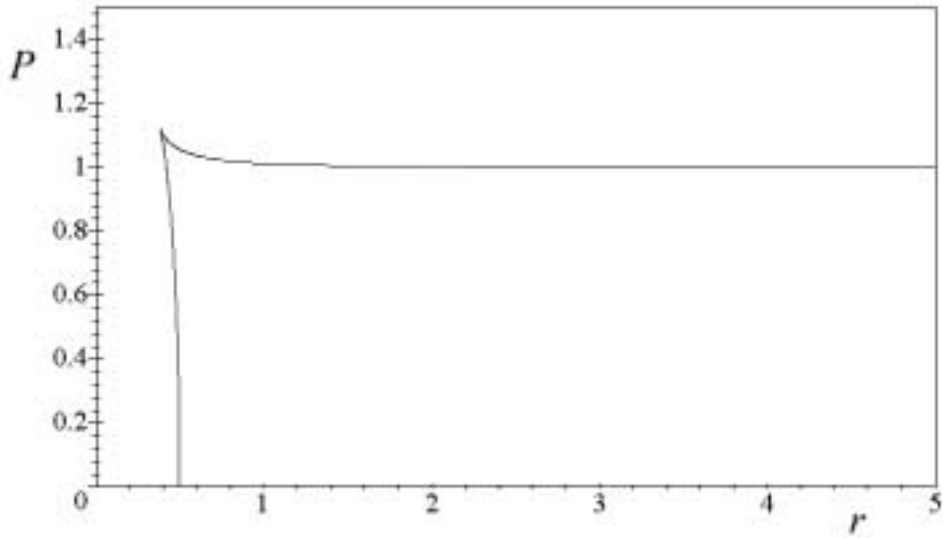

Fig.5 $P_{\text{CP1}}(r)$ relation described parametrically by $P(\xi)$ and $r(\xi)$ functions shown in Fig. 3.

As seen in the figure, the $P_{\text{CP1}}(r)$ relation contains two branches. Which of them presents the physically sensible solution we may find out checking within which range of $\xi$ the function of the square distance

$$f(\xi + \Delta\xi) = 2r^2 - 2r^2\cos(\xi + \Delta\xi) + \left(\frac{P}{2\pi}\right)^2(\xi + \Delta\xi)^2 \tag{9}$$

in which $P$ and $r$ are in the relation described by (8) displays a minimum versus $\Delta\xi$. (When the minimum exists, the turn-to-turn distance we are calculating becomes identical with the *doubly critical self distance* introduced by Simon [8]: $dcsd(h) = \min_{x \neq y}\left\{ |h(x) - h(y)| : h'(x) \perp (h(x) - h(y)),\ h'(y) \perp (h(x) - h(y)) \right\}$, where $h$ is the helix parameterised by arc-length, $h(x)$ and $h(y)$ are points located on the helix, $(h(x)-h(y))$ is the vector which joins the points and $h'(x)$ and $h'(y)$ are the tangent vectors at $h(x)$ and $h(y)$, respectively.) To reach the aim, we substitute (8) into (9), expand it into a Taylor series truncated at the $(\Delta\xi)^2$ term and differentiate it twice versus $\Delta\xi$. The second derivative obtained in such a manner equals:

$$\frac{d^2 f}{d(\Delta\xi)^2} = \frac{2\xi\cos(\xi) - 2\sin(\xi)}{\xi[\xi\sin(\xi) + 2\cos(\xi) - 2]}. \tag{10}$$

Fig. 6 presents it within the interesting $(\pi,\ 2\pi)$ range.

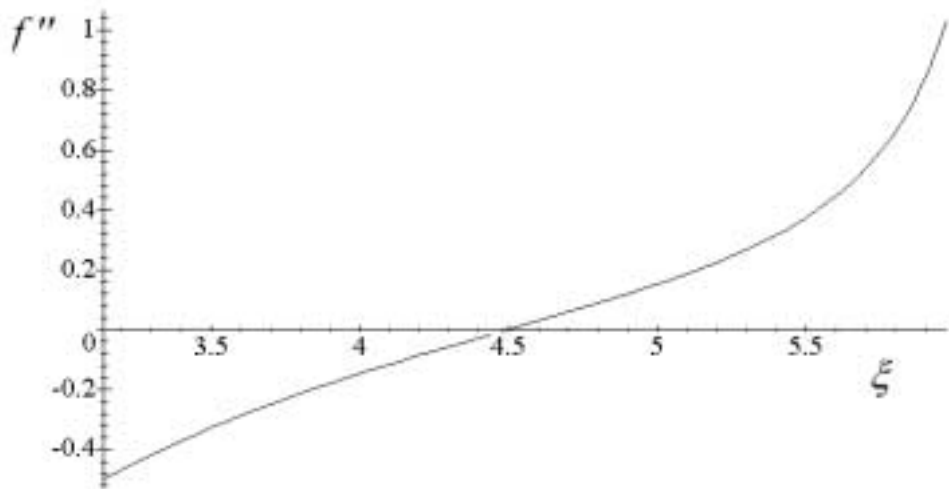

Fig. 6 The second derivative of the square distance function.

The second derivative is positive in the interval (4.49341, $2\pi$). Replotting the $P_{\mathrm{CP1}}(r)$ relation only within this range reveals which of its branches presents the required solution. See Fig. 7.

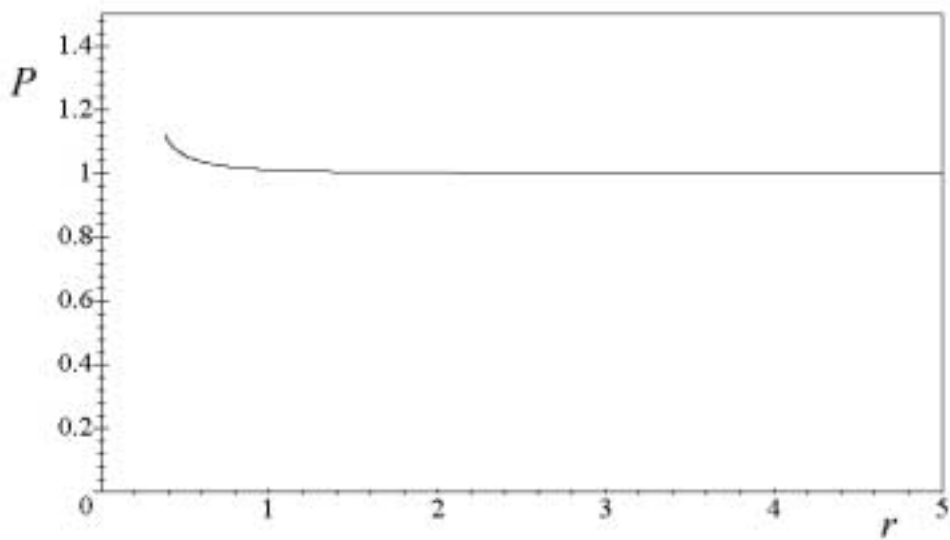

Fig.7 The dependence of the pitch $P$ on the radius $r$ found in closely packed helices. $D$=1.

The result we obtained stays in agreement with the intuition: a closely packed helix, whose radius is squeezed, increases its pitch. As seen in the figure, $r$=0.5 seems to limit the squeezing process. Is it really the case?

### 4. Closely packed helix limited by its curvature

There exists another mechanism, which limits the set of possible ($r,P$) values of the helices formed on an ideal rope. It stems from the fact that the ideal rope of diameter $D$ cannot have a local curvature larger than $2/D$. The following heuristic reasoning indicates the source of the limitation. Let $h(x)$ be a helix of curvature $\kappa$ parameterised by arc-length $x$. Let $h'(x)$ be the field of its tangent vectors. Imagine that a disk of diameter $D$, centred on $h(x)$ and perpendicular to $h'(x)$ is swept along the helix. The circular border of the moving disk determines within the space the surface of the ideal rope. Consider the traces $d(x_1)$ and $d(x_2)$ of the disk in two consecutive positions $h(x_1)$ and $h(x_2)$ separated by an infinitesimal arc $dx$. Because of the non-zero curvature of the helix along which the disk was swept disks $d(x_1)$ and $d(x_2)$ are not parallel to each other – they are inclined by angle $\kappa dx$. When $\kappa > 2/D$ the disks overlap. Consequently, the surface of the rope determined by border of the swept disk becomes non-smooth. Fig.8 illustrates the situation.

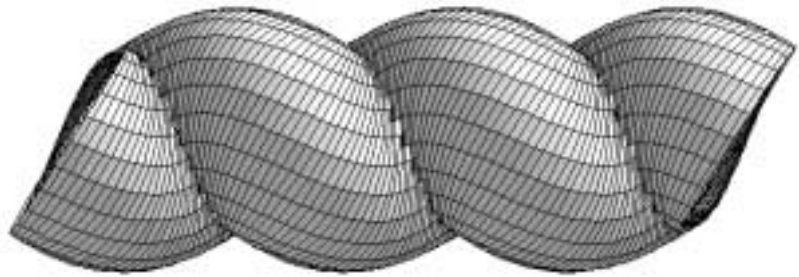

Fig. 8 If the local curvature of the helix is too large the impenetrability of the ideal rope is also violated. For the helix presented in the figure $r$=0.1and $P$=1, thus its local curvature $\kappa$=2.83.

Let us consider analytical consequences of this inequality.

The curvature of a helix defined by equation (1) equals:

$$\kappa = \frac{r}{r^2 + \frac{P^2}{4\pi^2}} \qquad (12)$$

As easy to find, equation $\kappa$=2/$D$ is fulfilled if

$$P_{CP2} = \pi\sqrt{2rD - 4r^2} \qquad (13)$$

Fig. 9 presents relation (13) together with the discussed above relation (8). One can see that their plots cross. As a result some parts of the $P_{\text{CP1}}(r)$ and $P_{\text{CP2}}(r)$ solutions become inaccessible.

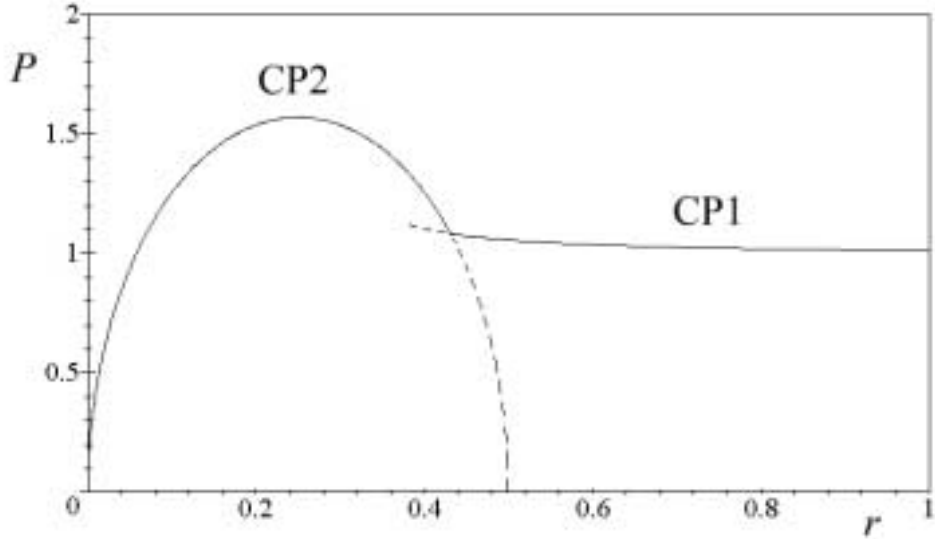

Fig.9 $P_{\text{CP1}}(r)$ and $P_{\text{CP2}}(r)$ solutions plotted together. The mutually inaccessible parts of the solutions are marked with a dashed line. $D$=1.

Numerically determined coordinates of the crossing point are as follows:

$$r_0 = 0.431092$$
$$P_0 = 1.08292 \qquad (14)$$

Mutually accessible parts of the $P_{\text{CP1}}(r)$ and $P_{\text{CP2}}(r)$ solutions define within the $(r,P)$ plane a border line $P_{\text{CP}}(r)$ below which one cannot go; helices found in this forbidden region are impossible to build with the perfect rope. Figure 10 presents the border $P_{\text{CP}}(r)$ line together with the images of the closely packed helices located at a few representative points of it .

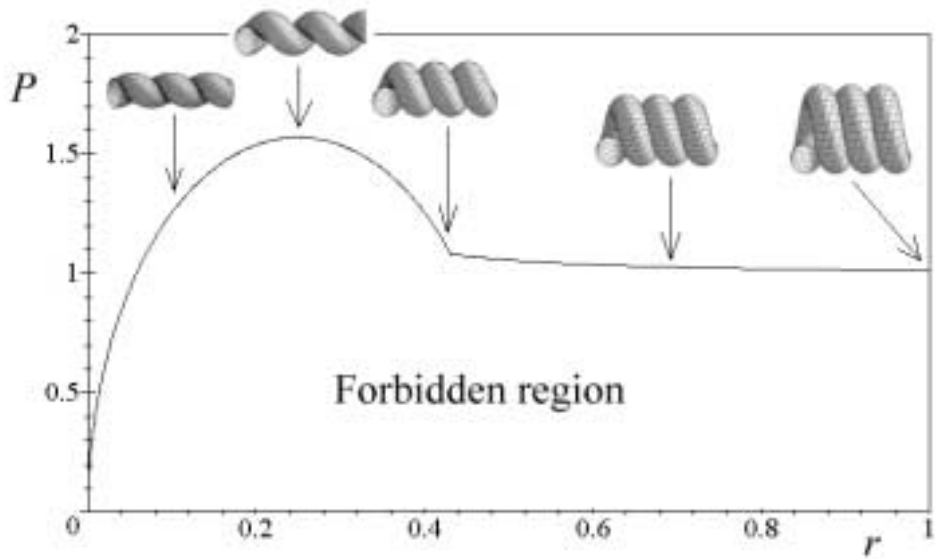

Fig. 10  Plot of the relation between the radius and the pitch of the closely packed helices together with their images at a few representative points of the plot.  *D*=1.

The helix seen in the cusp point at which the CP1 and CP2 solutions meet, is the optimal helix discussed in ref.[4].

## 5. Physical properties of the optimal helix

Monte Carlo simulations performed by Maritan et al. were aimed to find those shapes of the rope, for which the radius of gyration becomes minimised. Radius of gyration is a geometrical property of a curve. It is defined  as the root mean square distance of a set of points (obtained by a discretisation of the curve) from its centre of mass. Does the optimal helix minimises some physical properties of the curve?

Let us imagine that equal masses,  *m*=1, are distributed along an ideal rope at equal distances *dL*. The rope is then shaped into a closely packed helix. Which is the energy of the gravitational interaction between the masses? Obviously, the energy depends  on the Euclidean distance *dS* between them. See Fig. 11. The distance *dS* is always smaller than *dL* and its value depends on the parameters of the helix into which the rope is shaped. As shown above, values of the parameters are limited by the turn-to-turn distance (doubly critical self distance) and the curvature of the helix.

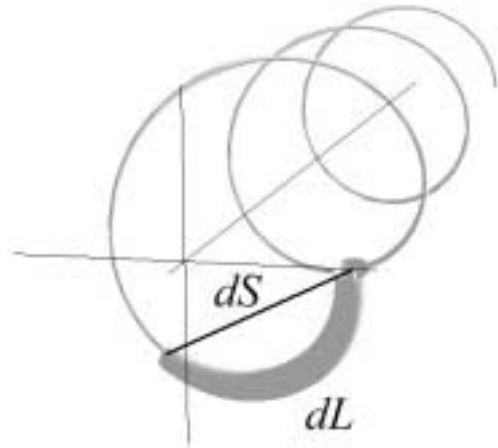

Fig. 11 The relation between *dL* and *dS*.

Calculations we performed show, that for a given *dL*, *dS* reaches its minimum within the optimal helix. See Fig. 12.

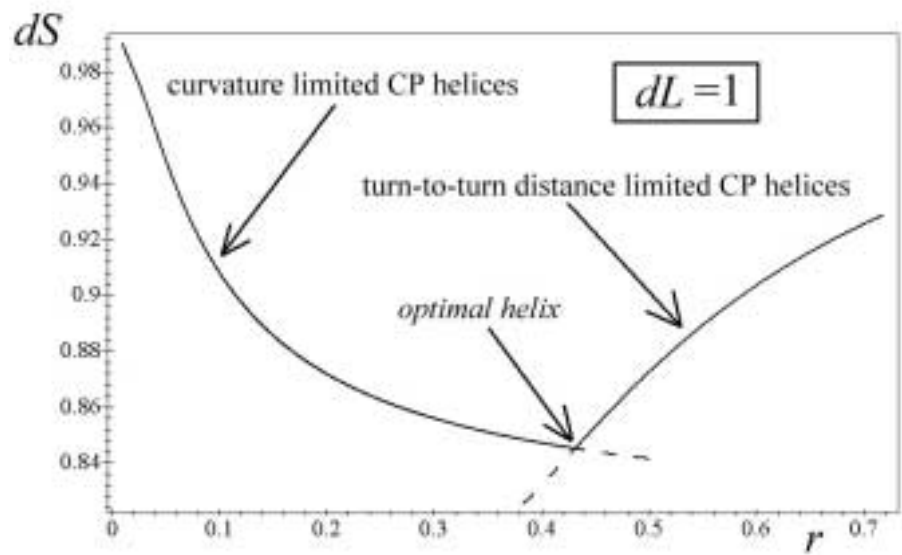

Fig. 12 *dS* vs. *r* for closely packed helices.

Consequently, the energy of gravitational interactions reaches its minimum as well. One can certainly find a few other cases, in which the optimal helix of Maritan et al. proves also to be optimal.

## 6. Discussion

We have shown that looking for closely packed helices formed with the ideal rope one has to consider two cases: helices limited by the turn-to-turn distance and helices limited by the local curvature. As indicated by Maritan et al. [4], the two cases can be brought into one: helices limited by the *global curvature*, a notion introduced by O. Gonzalez and J. Maddocks [9]. The radius of the global curvature at a given point $P$ of a space curve $C$ is defined as the radius of smallest circle which goes through the chosen point and any other two points $P_1$, $P_2$ belonging to $C$ and different from $P$. Putting a limit on the global curvature of a helix one limits both the local curvature and the closest distance between its consecutive turns. As a result the union of accessible parts of the presented above partial solutions $P_{CP1}(r)$ and $P_{CP2}(r)$ can be seen as a single solution $P_{CP}(r)$. The solution shown in Fig.10 answers the problem formulated as follows: *which is the relation between the pitch P and the radius of helices whose global curvature equals 2/D?* Asking simpler, synthetic questions helps finding simpler, synthetic answers.

There exists another, equivalent formulation of the problem. Instead of the global curvature one can use the notion of the *injectivity radius* [10]. The injectivity radius of a smooth curve $K$ is the maximum radius of the disks which centred on each point of $K$ and perpendicular to its tangent remain disjoint. In terms of the injectivity radius the closely packed helices can be seen as *the helices whose injectivity radius is equal to the predetermined radius of the used tube*.

PP thanks Andrzej Stasiak and Giovanni Dietler for helpful discussions. He also thanks Herbette Foundation, Institute of Experimental Physics at UNIL and the Mathematics Department at EPFL for financial supports.